\documentclass[useAMS,usenatbib]{mn2e}
\usepackage{amssymb}
\usepackage{epsf}
\usepackage{bm}

\title[Breaking stress]{Breaking stress of neutron star crust}
\author[A. I. Chugunov, C. J. Horowitz]
{A.~I.~Chugunov$^1$\thanks{andr.astro@mail.ioffe.ru},
C.~J.~Horowitz$^2$\\
$^1$Ioffe Physical-Technical Institute of the Russian Academy of
Sciences, Politekhnicheskaya 26, 194021 Saint-Petersburg, Russia
\\
$^2$Department of Physics and Nuclear Theory Center, Indiana
University, Bloomington, IN 47405}

\begin{document}

\date{Accepted 2010 xxxx. Received 2010 xxxx;
in original form 2010 xxxx}

\pagerange{\pageref{firstpage}--\pageref{lastpage}} \pubyear{2010}

\maketitle

\label{firstpage}

\begin{abstract}
The breaking stress (the maximum of the stress-strain curve) of neutron
star crust is important for neutron star physics including pulsar
glitches, emission of gravitational waves from static mountains, and
flares from star quakes.  We perform many molecular dynamic
simulations of the breaking stress at different coupling parameters
(inverse temperatures) and strain rates.  We describe our results with the
Zhurkov model of strength.
We apply this model to estimate the breaking stress for timescales
$\sim 1$~s -- 1 year, which are most important for applications, but
much longer than can be directly simulated.  At these timescales the
breaking stress depends strongly on the temperature.  For coupling
parameter $\Gamma\lesssim 200$ matter breaks at very small stress,
if it is applied for a few years.  This viscoelastic creep can limit
the lifetime of mountains on neutron stars.  We also suggest an
alternative model of timescale-independent breaking stress, which
can be used to estimate an upper limit on the breaking stress.
\end{abstract}


\maketitle
\section{Introduction}
Ions in the neutron star crust can form a Coulomb crystal which
determines the crust's elastic properties.  For example, the
breaking stress $\sigma_\mathrm{b}$ is the maximum stress, as a
function of strain, that the crust can support.  If larger stress is
applied to matter it will not remain in a static configuration.
Neutron star crust is under high pressure.  As a result, voids or
fractures do not occur (\citealt{HK09}) and this  simplifies how the
crust breaks.
The crust will break if the local stress is larger then the
breaking stress at at least one point in the crust. In this paper we
mostly refer to breaking stress (force per unit area).  The breaking strain (corresponding
value of the fractional deformation) can be estimated from the linear
stress-strain relation, where the slope is given by the well known elastic constant or shear modulus (\citealt{SOIIV91,HH08}).

There are a lot of models which associate breaking of neutron star
crust with observational phenomena. First of all, crust breaking
causes pulsar glitches in the starquake  model of
\cite{Ruderman69,Ruderman91}
%
%
and in some recent models (for brief review see \citealt{CH08}).
%
%
Second, some models of magnetar giant flares involve crust
breaking (\citealt*{Thompson01}).  Here the crust may need to be very
strong
to control the release of large magnetic energies responsible for
extremely energetic gamma ray bursts.  Indeed, our results, see
below, predict that the crust is very strong.  Finally, the breaking
stress limits the maximum size of ``mountains''\ on neutron stars.
Mountains, on rapidly rotating stars, can efficiently emit
gravitational waves (\citealt*{Ushomirsky00,Haskell07}) that could
be detected by present large scale interferometers
(\citealt*{Abbott07, Abbott08}).  Gravitational wave emission could
be especially important for low mass neutron stars that can have
large deformations (\citealt{Horowitz09}). Moreover the balance of
angular momentum gained from accretion, and radiated in
gravitational waves, can control the spin period of some accreting
stars (\citealt*{Watts08}).

For most terrestrial materials the durability (time before
breaking) $\tau$ at a given stress is known to depend on
the temperature and, of course, the applied stress (e.g.,
\citealt*{Zhurkov57,RST72,Slutsker05,Slutsker07}). In other words,
the breaking stress is not just a constant, defined by the matter
parameters (density, temperature and composition), but depends on the
duration of the stress -- the matter can break at lower stress, if one waits long enough.
In fact, the dependence of $\sigma_\mathrm{b}(\tau)$ on $\tau$ is logarithmic and we
will refer to $\tau$ as the timescale of the process. The aim of the
present paper is to determine how $\sigma_\mathrm{b}$, depends on $\tau$
and on temperature $T$ for neutron star crust.   We suggest a simple
expression [see Eq.\ (\ref{Zhurkov_dimless}) parameterized by Eq.\
(\ref{ZhurFit}) below] for the durability of neutron star crust
material at a given stress.  This expression gives the lifetime of elastic
deformations (mountains, for example) and one can easily extract
the breaking stress for the timescale of the process of interest (a few
years for pulsar glitches, for example).

We also note the possible application of our results to the physics
of dusty plasmas.  Such a plasma can form a Yukawa crystal, just
like  the ions in neutron star crust.  Its breaking stress can be
measured in the laboratory as a threshold stress which is needed
to obtain flow. The presence of such a threshold has been recently
experimentally demonstrated by \cite{Gavrikov09}.

Until recently breaking stress studies
(\citealt{Smoluchowski70,Ruderman91}) have been based on analogies
with terrestrial materials instead of accurate calculations, and the
uncertainties were large.  Recently, large scale molecular dynamics
(MD) simulations of  the breaking stress were performed
(\citealt{HK09}). The breaking strain was found to be large $\sim
0.1$ and only slightly affected by impurities, defects, and
polycrystalline structure (grain boundaries).  In the present paper,
we extend these results with extensive MD simulations of breaking
stress at different temperatures and strain rates.  We use these
results to extract parameters of the Zhurkov model of strength [see
Sec.\ \ref{Sec:Zhurkov} and \cite{Zhurkov57,RST72} for details] and
apply this model to estimate the breaking stress for timescales of 1
s -- 1 year, which are of most interest for applications, but much
longer than possible for direct MD simulations.

\section{Formalism}\label{Sec:Fomalizm}

The neutron star outer crust consist of ions and electrons. The
inner crust also contains free neutrons (see e.g.
\citealt*{BOOK,CH08}). The electrons are degenerate and form a
slightly polarizing background which screens the Coulomb interaction
between ions. We describe electron screening by a Thomas-Fermi
screening length
\begin{equation}
 \lambda_\mathrm{e}=\frac{\sqrt{\pi}}{2}
 \left(\frac{\hbar c}{e^2}\right)^{1/2}
 \left(3\pi^2 n_\mathrm{e}\right)^{-1/3},
\end{equation}
 where
$n_\mathrm{e}$ is the electron number density, $e$ is elementary
charge and $c$ is the speed of light.  For simplicity, we assume the
electrons are ultrarelativistic.

In this paper we discuss a one component plasma, where all ions have
the same charge number $Z$. The ions interact via a Yukawa potential
\begin{equation}
\phi(r)=\frac{Z^2e^2}{r}\,\exp\left(-r/\lambda_\mathrm{e}\right),
\end{equation}
where $r$ is the interion distance.   The state of the ion system
can be characterized by the classical coupling parameter
%
\begin{equation}
  \Gamma=\frac{Z^2 e^2}{aT}.
\end{equation}
Here $a=\left[3/(4\pi n_\mathrm{i})\right]^{1/3}$ is the ion sphere
radius and $T$ is the temperature in energy units. The ion number
density is $n_\mathrm{i}=n_\mathrm{e}/Z$ because the system is
neutral.

In the ultrarelativistic limit, the ratio of the electron screening
length $\lambda_\mathrm{e}$ to the ion sphere radius $a$ depends
only on $Z$: $\lambda_\mathrm{e}/a\approx 5.41/Z^{1/3}$. We use
$Z=29.4$ in all of our calculations. This is the mean charge of the
ions in the crust composition used by \cite*{HBB07} and \cite{HK09}.
It corresponds to $\lambda_\mathrm{e}/a\approx 1.75$. For such
screening length the crystal is thermodynamically stable at
$\Gamma>\Gamma_\mathrm{m}\approx 176.1\pm0.7$ (\citealt{HBB07}),
which is close to the melting point $\Gamma_\mathrm{m}=175$ expected for the one component plasma in the absence of electron screening (\citealt{PC00}). We suppose, that the breaking stress depends only slightly on screening length and our results can
be used for matter of any composition.

A typical dynamical frequency is the ion plasma frequency $\omega_\mathrm{p}$,
\begin{equation}
 \omega_\mathrm{p}=\sqrt{\frac{4\pi Z^2 e^2 n_\mathrm{i}}{m_\mathrm{i}}},
\end{equation}
where $m_\mathrm{i}=A m_\mathrm{u}$ is the ion mass. The ratio of
plasma temperature $T_\mathrm{p}=\hbar \omega_p$ to the temperature
$T$ characterizes the importance of quantum effects on ion motion.
In our classical MD simulations we neglect quantum effects (assuming
$\hbar\rightarrow 0$), but they can be important at the conditions
of realistic neutron star crust and we discuss them in Sec.\
\ref{Sec:ZeroPoint}.

\subsection{Molecular dynamics simulations}

To calculate the breaking stress we developed a parallel version of
the YUKAWAMD code (\citealt{HK09}) where the ion system is strained
by deforming periodic boundaries.  We use the velocity Verlet
algorithm (\citealt*{MDVerlet}) with a time step $\sim
0.1/\omega_\mathrm{p}$. The simulations are performed as follows:
the original system is taken to be a perfect body centered cubic
crystal contained in a cubic box, that is aligned with the lattice
planes. After thermalization, we start a shearing deformation of the
periodic box boundaries according to $x\rightarrow x+\epsilon y/2$,
$y\rightarrow y+\epsilon x/2$, and $z\rightarrow
z/(1-\epsilon^2/4)$. This deformation conserve volume of the
simulation box. Here the strain $\epsilon=vt$ increases with time at
constant strain velocity $v$. Since the deformation is adiabatic, we
define stress as $\sigma=\partial \mathcal{E}/\partial \epsilon$
(\citealt{Landau7}), where $\mathcal{E}$ is internal energy at unit
volume. The derivative is calculated numerically as
$\sigma(\epsilon)=\left[\mathcal{E}(\epsilon)
-\mathcal{E}(\epsilon-\Delta\epsilon)\right] /\Delta\epsilon$ with
$\Delta\epsilon=0.002$. The breaking stress $\sigma_\mathrm{b}$ is
the maximum stress obtained during the shear simulation.
The corresponding strain we refer to as the breaking strain
$\epsilon_\mathrm{b}$.

To decrease the simulation time we prepare a well thermalized system
at strain $\epsilon=0.05$ and use it as an initial configuration for
deformations with different $v$. For the longest runs (with lowest
$v$) we start from a configuration at larger strain. The initial
parameters of this configuration are extracted from other simulations
with larger $v$.
We also use a cutoff radius $r_\mathrm{cut}$ to omit interactions at
$r>r_\mathrm{cut}$ and increase computation speed.  Most of
the simulations are done for $9826$ ions in a periodic box and
$r_\mathrm{cut}=13 \lambda_\mathrm{e}$. This large $r_\mathrm{cut}$
value is needed to accurately reproduce the shear modulus (see
\citealt{HH08}).   We discuss these simulation parameters in more
detail in Sec.\ \ref{Sec:Results}.

\subsection{Zhurkov model of strength}\label{Sec:Zhurkov}

The Zhurkov model (see e.g. \citealt{Zhurkov57,RST72,Slutsker05}) is
based on the kinetic conception of strength. The main assumption is
the following. The breaking event occurs due to thermal
fluctuations. To achieve breaking, a fluctuation should have
energy $U$.  This energy threshold is reduced by the product of the
stress $\sigma$ times an activation volume $V$. The probability of
a breaking fluctuation is $\propto \exp\left[-(U-\sigma V)/T\right]$.
Following this model, the matter will break at stress
$\sigma_\mathrm{b}$, if it is applied over the time interval $\tau$
\begin{equation}\label{Zhurkov}
     \tau=\tau_0\exp\left(\frac{U-\sigma_\mathrm{b} V}{T}\right),
\end{equation}
where $\tau_0$ is a typical timescale for ion motion, that we take
to be $\tau_0=\omega_\mathrm{p}^{-1}$.  Defining dimensionless
threshold energy $\bar{U}=U\,a/(Z^2e^2)$  and stress
$\bar\sigma=\sigma/(n_\mathrm{i} Z^2e^2/a)$, Eq.\ (\ref{Zhurkov})
can be rewritten in the form
\begin{equation}\label{Zhurkov_dimless}
     \tau=\frac{1}{\omega_\mathrm{p}}
     \exp\left(\bar{U}\Gamma
     -\bar\sigma_\mathrm{b} \bar{N} \Gamma\right),
\end{equation}
where $\bar{N}=Vn_\mathrm{i}$ is the number of ions in the
activation volume.  We use $\bar{U}$ and $\bar{N}$ as fit parameters
to reproduce MD data. Equation\ (\ref{Zhurkov_dimless}) is written
for static stress, but our MD simulations are dynamical.  The strain
is linearly increasing with time, this produces a stress that also increases, almost linearly, with time  $\sigma\approx\mu\epsilon=\mu vt$.  Of course,
the linear dependence of $\sigma$ on $t$ is not correct after the breaking
event.  Here $\mu\approx\sigma_\mathrm{b}/\epsilon_\mathrm{b}$ is the
corresponding elastic constant.  Equation\ (\ref{Zhurkov_dimless}) can
be easily generalized for such deformation (see \cite{Slutsker83},
for the example). The strain velocity $v$ and breaking stress are
related by the equation
\begin{equation}
 \frac{v}{\omega_\mathrm{p}}=
 \frac{\epsilon_\mathrm{b}}{\bar{N}\Gamma\bar\sigma_\mathrm{b}}
    \exp\left(-\bar{U}\Gamma+\bar\sigma_\mathrm{b} \bar{N}\Gamma\right).
    \label{Zhurkov_dyn}
\end{equation}
The stress is lower than $\sigma_\mathrm{b}$ during most of the
deformation time.  As a result, the time before breaking
$\tau=\epsilon_\mathrm{b}/v$ is larger than $\tau$ given by Eq.\
(\ref{Zhurkov_dimless}) by a factor of
$\bar{N}\Gamma\bar\sigma_\mathrm{b}$.

\section{Results}\label{Sec:Results}

\begin{figure}\begin{center}\leavevmode
 \epsfxsize=3.2in
    \epsfbox{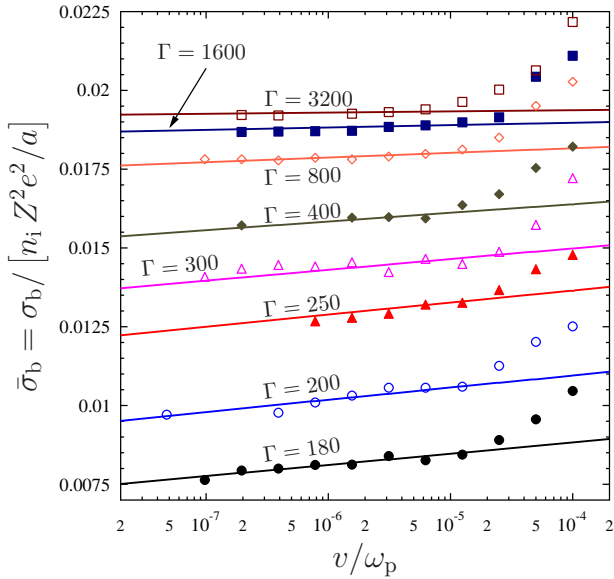}
    \end{center}
    \caption{Dimensionless breaking stress
    $\bar\sigma_\mathrm{b}$
    versus dimensionless shear speed $v/\omega_\mathrm{p}$.
    The symbols are MD data (dots, circles, filled triangles, triangles,
    filled rhombus, rhombus, filled squares and squares
    are for $\Gamma=180$, 200, 250, 300, 400, 800, 1600 and 3200 respectively).
    The lines are predictions of the Zhurkov model,
    Eq.\ (\ref{Zhurkov_dyn}), with parameters given by Eq.\
    (\ref{ZhurFit}). }
    \label{Fig:Breaking9826}
\end{figure}
Our MD simulation results for the breaking stress of a  system of
$9826$ ions are shown by symbols in Fig.\ \ref{Fig:Breaking9826}.
The data are fitted by Eq.\ (\ref{Zhurkov_dyn}) with parameters
$\bar U$ and $\bar N(\Gamma)$,
\begin{equation}
   \bar{U}=0.366,\quad \bar{N}=\frac{500}{\Gamma-149}+18.5.
   \label{ZhurFit}
\end{equation}
This fit is shown as lines in Fig.\ \ref{Fig:Breaking9826}. The
$\epsilon_\mathrm{b}$ in Eq.\ (\ref{Zhurkov_dyn}) was taken from
spline interpolation of MD data over $v/\omega_\mathrm{p}$. In fact,
$\bar\sigma_\mathrm{b}$ depends on $\epsilon_\mathrm{b}$ only
logarithmical and one can set
$\epsilon_\mathrm{b}=\epsilon_\mathrm{b}(\Gamma)$ (and neglect
the dependence on $v/\omega_\mathrm{p}$) in Eq.\ (\ref{Zhurkov_dyn}) to
reproduce the plot. There is good agreement between the fit and MD
results, except for $v/\omega_\mathrm{p}\gtrsim
2\times10^{-5}$.
We suppose, that here the deformation velocity is too rapid and the
breaking stress increases because there is not enough
time for the system to break and relax the lattice.  We exclude these points from
the fit.

If Eqs.\ (\ref{Zhurkov_dimless}) and (\ref{ZhurFit}) are formally
applied for $\Gamma$  below the melting value $\Gamma_\mathrm{m}$,
we find that the crystal has a non-vanishing durability at fixed stress
until $\Gamma\gtrsim 150$. This is in qualitative agreement with the
results of \cite{Daligault06}, who found Coulomb crystals to be
metastable for $\Gamma\gtrsim 150$.

\subsection{Dependence on system size and cutoff radius}

To study finite size effects we have performed a simulation
with $250000$ ions at $\Gamma=800$.  For a strain speed
$v/\omega_\mathrm{p}=6.25\times 10^{-6}$ we obtain a breaking stress
$\bar\sigma_\mathrm{b}=0.0187$. This is only 4\% larger than the
value $0.0180$ obtained for a $9826$ ion simulation. We
conclude that finite size effects are not very large.

To study the dependence on cutoff radius $r_\mathrm{cut}$ we perform
a set of additional runs with $r_\mathrm{cut}=10\lambda_\mathrm{e}$
and $r_\mathrm{cut}=16\lambda_\mathrm{e}$. The value
$r_\mathrm{cut}=10\lambda_\mathrm{e}$ is too small (for $\Gamma=800$
the breaking stresses are systematically 10\% larger than for
$r_\mathrm{cut}=13\lambda_\mathrm{e}$), but results for
$r_\mathrm{cut}=13\lambda_\mathrm{e}$ and
$r_\mathrm{cut}=16\lambda_\mathrm{e}$ are almost the same (within a
few percent statistical accuracy).
We conclude that $r_\mathrm{cut}=13\lambda_\mathrm{e}$ is large
enough.  It is easy to understand why so large a value of
$r_\mathrm{cut}$ is needed.  Let us estimate the interaction energy
$U_\mathrm{cut}$ of an ion with other ions at distances
$r>r_\mathrm{cut}$.  Assuming the ions are uniformly distributed for
$r>r_\mathrm{cut}$ one obtains
\begin{equation}
    U_\mathrm{cut}= 3\frac{Z^2 e^2}{a}
    \frac{\lambda_\mathrm{e}\,(r_\mathrm{cut}+\lambda_\mathrm{e})}{a^2}
    \exp(-r_\mathrm{cut}/\lambda_\mathrm{e}).
\end{equation}
For $r_\mathrm{cut}=10\lambda_\mathrm{e}$ and
$\lambda_\mathrm{e}=1.75 a$ we obtain $U_\mathrm{cut}\approx Z^2
e^2/220a$ which is small compared with the Coulomb energy, but for
$\Gamma\gtrsim \Gamma_\mathrm{m}\approx176$ it is of the same order
of magnitude as the thermal energy, which is crucial for the
breaking of the crystal.  For $r_\mathrm{cut}=13\lambda_\mathrm{e}$
the energy $U_\mathrm{cut}\approx Z^2 e^2/3400a$, which is small
compared with thermal energy up to $\Gamma\lesssim 1600$.

\subsection{Correction for quantum effects}\label{Sec:ZeroPoint}

In our classical MD simulations we can not properly include quantum
effects, such as zeropoint vibrations.  But such effects can be
important for applications, since the crust temperature can be less
than the plasma temperature. At $T\ll T_\mathrm{p}$ the breaking
event takes place because of subbarrier tunneling, but not
overbarrier thermal fluctuations as in a classical crystal (see
\citealt{Slutsker83}, for example). To include quantum effects we
make a simple assumption that the breaking stress mainly depends on
the root-mean-square displacement of the ions.  A similar idea was
suggested by \cite{Salganik70} to describe polymers breaking and the
results were shown to be in a good agreement with experiments with
boron samples (\citealt{Slutsker83}).  Let us introduce a
renormalized coupling parameter $\tilde{\Gamma}$ in such a way, that
the classical ion crystal at $\Gamma=\tilde{\Gamma}$ has
approximately the same root-mean-square displacement as the quantum
system at a given $\Gamma$ and $T/T_\mathrm{p}$
\begin{equation}
\tilde{\Gamma}=\Gamma\,\left[1+\frac{1}{4}\,\frac{T^2_\mathrm{p}\,u^2_{-1}
}{T^2\,u_{-2}^2}\right]^{-1/2} \approx
\Gamma\,\left[1+0.013\,\frac{T_\mathrm{p}^2}{T^2}\right]^{-1/2}.
\label{quantum}
\end{equation}
Here $u_{-1}\approx 2.7986$ and $u_{-2}\approx 12.973$ are moments
of the phonon spectrum (see \citealt{BPY01}, and the values given
are for a body centered cubic crystal).  We assume that
$\Gamma=\tilde{\Gamma}$ substituted in Eqs.\ (\ref{Zhurkov_dimless})
and (\ref{ZhurFit}) provide qualitatively correct results not only
for classical crystals ($T\gg T_\mathrm{p}$) but also for quantum
crystals. In Sec.\ \ref{Sec:Discus} it will be shown that the quantum
corrections are typically not very large.

\section{Discussion}\label{Sec:Discus}

\begin{figure}
    \begin{center}
        \leavevmode
        \epsfxsize=3.2in \epsfbox{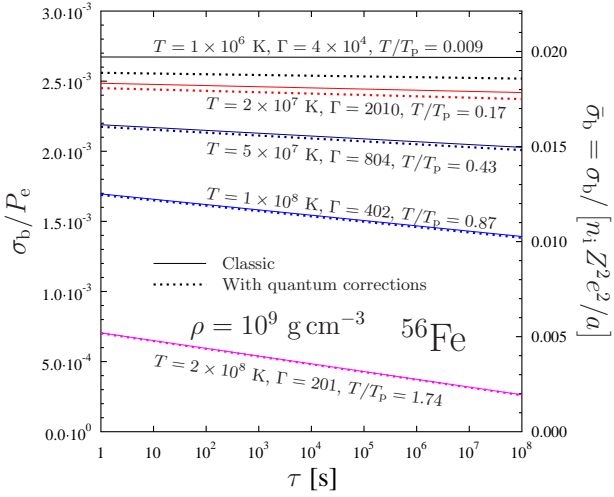}
    \end{center}
    \caption{The ratio of the breaking stress $\sigma_\mathrm{b}$ to
    the electron pressure $P_\mathrm{e}\approx 4.4\times 10^{26}$~dyn\,cm$^{-2}$ (left axis)
    versus timescale of the stress $\tau$ at five temperatures $T=1
    \times 10^6$, $2\times 10^7$, $5\times 10^7$, $1\times 10^8$ and $2
    \times 10^8$~K for $^{56}$Fe matter at the density
    $10^9$~g\,cm$^{-3}$.
    The solid lines are predictions of the Zhurkov model,
    Eq.\ (\ref{Zhurkov_dyn}), with parameters given by Eq.\
    (\ref{ZhurFit}) without corrections for quantum effects.
    The dotted lines are corrected for quantum effects in accordance
    with Sec.\ \ref{Sec:ZeroPoint}.
    }
    \label{Fig:BreakExample}
\end{figure}
The knowledge of the parameters [Eq.\ (\ref{ZhurFit})] of the
Zhurkov model of strength [Eq.\ (\ref{Zhurkov_dimless})] allow us
to estimate the breaking stress for timescales $\sim 1$ s to 1 year
which are the most interesting for astrophysical applications.
Direct MD simulations for such timescales are impossible because of
the very small dynamical timescale of the ions
$\sim\omega_\mathrm{p}^{-1}\sim 10^{-20}$~s.  One would need at
least a few$\times10^{20}$ MD steps to simulate one second, but one
time step takes approximately 10 core-seconds of computer time.

Our estimates of the long time breaking stresses are shown in Fig.\
\ref{Fig:BreakExample} for $^{56}$Fe matter at a density
$10^9$~g\,cm$^{-3}$.  For each of 5 temperatures ($T=1 \times 10^6$,
$2\times 10^7$, $5\times 10^7$, $1\times 10^8$ and $2\times 10^8$~K)
the figure contains two lines: the solid line corresponds to the
Zhurkov model for classical crystals [Eqs.\ (\ref{Zhurkov_dimless})
and (\ref{ZhurFit})] and the dotted line includes corrections for
quantum effects as described in Sec.\ \ref{Sec:ZeroPoint}.

One can see a strong dependence of the breaking stress on the
temperature. For the highest temperature $T=2\times 10^8$~K the
estimated breaking stress almost vanishes on the timescale of a few
years.  This can limit the lifetime of mountains, supported by
elasticity, on neutron stars with warm crust.   For lower
temperatures the breaking stress is significantly larger and the
$T$-dependence becomes weaker -- the breaking stresses at $T=10^8$
and $T=10^6$~K differ by less than a factor of two.  Finally, for
$T\lesssim 10^6$~K the breaking stress is almost independent of
temperature. The timescale dependence is strong for high temperatures
$T\sim 2\times 10^8$~K and almost vanishes for low temperatures
 $T\lesssim 2\times 10^7$~K.

In our model (Sec.\ \ref{Sec:ZeroPoint}) corrections for quantum
effects are not very large (compare solid and dotted lines in Fig.\
\ref{Fig:BreakExample}). The reason is simple.  At large
temperatures the ions are almost classical, and quantum corrections
are small. To describe low temperature limit let us estimate
breaking stress $\bar \sigma_\mathrm{b}\approx \bar U/N+\Delta\bar
\sigma_\mathrm{b}$, where correction $\Delta \bar
\sigma_\mathrm{b}\approx -\ln(\tau\omega_\mathrm{p})/(\Gamma\bar N)$
is small at low temperature. In our model the quantum corrections
are described by decreasing of the effective coupling parameter
$\tilde \Gamma$,
and affect only the correction therm $\Delta \bar
\sigma_\mathrm{b}$.

We should note, that our estimates for 1 s to 1 year timescales are
based on the extrapolation of the MD data of more than ten orders of
magnitude in time. However our data only span three orders of
magnitude in time ($10^{-7}\lesssim v/\omega_\mathrm{p}\lesssim
10^{-4}$). So the extrapolation results should be taken with
caution, especially for low coupling $\Gamma\lesssim 250$ where the
dependence of the breaking stress on time is most significant. For
large enough $\Gamma\gtrsim 800$ the breaking stress extrapolated to
timescales of a few years is only slightly lower than at timescales
achieved in our MD simulations (compare Fig.\ \ref{Fig:Breaking9826}
and the right axis of Fig.\ \ref{Fig:BreakExample}).  We expect that
the extrapolation for such strong coupling is more reliable.
However, we can not exclude that there are instabilities of the
deformed crystal which have large enough timescales to be
insignificant in our MD simulations but could reduce the breaking
stress for few second timescales. In any case, the breaking stress
at long timescales can not exceed the breaking stress
$\sigma_\mathrm{b}^\mathrm{max}$ obtained in our longest
simulations. The corresponding values can be fitted as
\begin{equation}
\sigma_\mathrm{b}^\mathrm{max}=\left(0.0195-\frac{1.27}{\Gamma-71}\right)\,
n_\mathrm{i}\frac{Z^2e^2}{a}. \label{sigma_max}
\end{equation}
This fit provides an upper limit for the long time breaking stress.

\section{Conclusions}\label{Sec:Concl}

We have performed extensive MD studies of the breaking stress of
neutron star crust.  Our results are in good agreement with the
Zhurkov model of strength, and we have  determined the corresponding
parameters, Eqs.\ (\ref{Zhurkov_dimless}) and (\ref{ZhurFit}).  We
apply this parametrization to estimate the breaking stress for very
long timescales $\sim 1$ s to 1 year, which are too large for direct
MD studies. We demonstrate that for a coupling parameter
$\Gamma\lesssim 200$ neutron star crust matter can break under very
small stress, if it is applied for a few years. This result is based
on an extrapolation of over ten orders of magnitude in timescale and
should be treated with caution.  We construct a qualitative model to
include the influence of quantum corrections on breaking stress
(Sec.\ \ref{Sec:ZeroPoint}).  We show that these corrections are
small (solid and dotted lines in Fig.\ \ref{Fig:BreakExample}). We
also present a fit for the upper limit on the breaking stress for
long times, Eq.\ (\ref{sigma_max}), which is based on the breaking
stresses for the slowest deformations directly obtained in our MD
simulations.  Our results can be used to estimate the lifetime of
mountains on neutron stars and to obtain the breaking stress for the
relevant timescales of other astrophysical phenomena associated with
crust breaking.

In this paper we concentrated on shear deformations of a perfect
body centered cubic crystal along the lattice planes.  Realistic
neutron star crust can contain impurities, grain boundaries and the
deformation may not be aligned with the lattice.  These  effects
were found to be unimportant by \cite{HK09}, but more detailed
studies are need.  In this paper we also considered only one ratio
of screening length to ion sphere radius
$\lambda_\mathrm{e}/a=1.75$, but this ratio depends on the
composition and can affect the breaking stress.  In addition,
breaking stress for the tension deformations can be important. 
%
We plan to study these effects in the near future.

\section*{Acknowledgments}
The authors are grateful to A.Y.~Potekhin, P.S.~Shternin,
A.I.~Slutsker and D.G.~Yakovlev for useful comments and to Don Berry
for help with parallel programming.

 This work
was partially supported by the Russian Foundation for Basic Research
(grant 08-02-00837), by the State Program ``Leading Scientific
Schools of Russian Federation'' (grant NSh 3769.2010.2), by the
President grant for young Russian scientists (MK-5857.2010.2), by
United States DOE grant (DE-FG02-87ER40365) and by Shared University
Research grants from IBM, Inc. to Indiana University.


\label{lastpage}
\end{document}